# Femtosecond charge and spin dynamics in CoPt alloys

Martin Pavelka[1], Simon Marotzke[2,3], Ru-Pan Wang[2], Mohamed F. Elhanoty[1], Günter Brenner[2], Siarhei Dziarzhytski[2], Somnath Jana[4], W. Dieter Engel[4], Clemens v. Korff Schmising[4], Deeksha Gupta[5], Igor Vaskivskyi[6], Tim Amrhein[7], Nele Thielemann-Kühn[7], Martin Weinelt[7], Ronny Knut[1], Juliane Rönsch-Schulenberg[2], Evgeny Schneidmiller[2], Christian Schüßler-Langeheine[5], Martin Beye[2,8], Niko Pontius[5], Oscar Grånäs[1], Hermann A. Dürr[1*]

[1] Department of Physics and Astronomy, Uppsala University, Box 516, 75120 Uppsala, Sweden

[2] Deutsches Elektronen-Synchrotron, DESY, Notkestraße 85, 22607 Hamburg, Germany

[3] Institut für Experimentelle und Angewandte Physik, Christian-Albrechts-Universität zu Kiel, Olshausenstraße 40, 24098, Kiel, Germany

[4] Max Born Institute for Nonlinear Optics and Short Pulse Spectroscopy, Max-Born Straße 2A, 12489 Berlin, Germany

[5] Helmholtz-Zentrum Berlin für Materialien und Energie, Albert-Einstein-Str. 15, 12489, Berlin, Germany

[6] Jozef-Stefan-Institute, Jamova cesta 39, 1000 Ljubljana, Slovenia

[7] Fachbereich Physik, Freie Universität Berlin, Arnimallee 14, 14195 Berlin, Germany

[8] Fysikum, Stockholm University, 10691 Stockholm, Sweden

*Corresponding Author E-mail: hermann.durr@physics.uu.se

ABSTRACT:

The use of advanced X-ray sources plays a key role in the study of dynamic processes in magnetically ordered materials. The progress in X-ray free electron lasers enables the direct and simultaneous observation of the femtosecond evolution of electron and spin systems through transient X-ray absorption spectroscopy (XAS) and X-ray magnetic circular dichroism (XMCD), respectively. Such experiments allow us to resolve the response seen in the population of the spin-split valence states upon optical excitation. Here, we utilize circularly polarized ultrashort soft X-ray pulses from the new helical afterburner undulator at the free-electron laser FLASH in Hamburg to study the femtosecond dynamics of a laser-excited CoPt alloy at the Co $L_3$ absorption edge. Despite employing a weaker electronic excitation level we find a comparable demagnetization for the Co $3d$-states in CoPt compared to previous measurements on CoPd. This is attributed to distinctly different orbital hybridization and spin-orbit coupling between $3d$ and $4d$ vs. 3d and $5d$ elements in the corresponding alloys and multilayers.

I. INTRODUCTION

Ferromagnetic alloys play an essential role in modern data storage technologies. The most typical elements to combine with transition metal ferromagnets (Co, Fe, and Ni) are heavier elements like Pd or Pt. Such a combination of $3d$ and $4d/5d$ elements increases the spin-orbit coupling and, in turn, often leads to perpendicular magnetic anisotropy that is favorable for magnetic recording. While such materials can be easily engineered and manufactured using modern sputter-deposition methods, the frontier of their application remains in manipulating their magnetization beyond the conventional field switching. An advanced manipulation route has been identified by using an intense femtosecond laser, obtaining ultrafast demagnetization in which the magnetization is reduced on a timescale shorter than one ps, initially reported on pure Ni [1]. Over the last three decades, this phenomenon has led to many new developments in fundamental and applied physics as well as instrumentation for the detailed study of the underlying microscopic mechanisms in a broad multitude of magnetic solids [2].

One of the keys to understanding ultrafast demagnetization in metallic ferromagnets is the role of hot electrons in the initial stage of the demagnetization process. An emerging topic in recent investigations has been the spin transfer mechanism in systems with multiple magnetic sublattices or elements [3-5]. So far, it is unclear whether spin transfer and the spin-flip mechanism are mediated via spin-orbit coupling or ultrafast magnon generation. This fundamental issue is tightly linked with the key application in all-optical magnetization switching [6], yet the underlying microscopic mechanism remains unexplained in ferromagnets [7]. Contrary to that, the mechanism has been explained in ferrimagnets, revealing the essential role of two magnetic sublattices allowing such switching to happen in alloys containing two elements with magnetic moments oriented in opposite directions [3].

Element-specific observations of ultrafast magnetization processes are now possible due to the development of advanced X-ray sources with a sub-picosecond pulse duration. The key techniques for the experiments mentioned above are time-resolved X-ray absorption spectroscopy (XAS) and X-ray magnetic circular dichroism (XMCD) measurements performed at L-edges in the soft-X-ray spectral range. When extended into the femtosecond regime, transient XMCD provides quantitative access to the temporal evolution of the spin and orbital magnetic moments [8-10]. Meanwhile, the demagnetization dynamics are accompanied by transient XAS changes that result in a spectral lineshape resembling a peak shift in addition to its reduction. This combination of transient XAS and XMCD in the femtosecond regime offers the possibility to observe the co-evolution of empty spin-polarized electronic states in the vicinity of the Fermi level after an optical excitation [11]. Such state-resolved experiments are promising for investigating the spin transfer mechanism, particularly in the transfer of spin-polarized electrons between 3$d$ transition metals and heavier 4$d$ and 5$d$ elements through their hybridized bands.

A prominent example for the necessary instrumental features is the femtoslicing capability at synchrotrons [8, 12], which offers the time-resolved observation of the demagnetization dynamics of 3$d$ transition metal and 4$f$ rare-earth elements. Unfortunately, femtoslicing sources typically reach a temporal resolution of 100 fs, which limits their use in probing the crucial timescales of the initial tens of femtoseconds. Observing these initial stages of demagnetization on the few femtosecond time-scale is, in general, one of the most significant experimental challenges in the field. The most promising and advanced experimental approach has emerged from the development of the X-ray free electron lasers (FELs) with the notable example of LCLS [13], where for the first time, an X-ray FEL facility operated a helical afterburner undulator for the production of X-ray photons with circular polarization, key for magnetization studies [14]. Recently, the FERMI FEL has also demonstrated the capability of generating soft X-ray photons with circular polarization [15]. As will be shown below, we employ the newly installed helical afterburner undulator at the FLASH FEL facility for the measurements reported here.

The experimental studies of X-ray interaction with materials during ultrafast demagnetization have been accompanied by various theoretical approaches for modeling the transient lineshape changes. In the initial femtosecond measurements performed at the Ni $L_3$ edge [8], the XAS change was attributed to electronic structure modifications [8, 10]. Further theoretical work ensued, with the prominent example [16] that the zero-crossing in the transient differential XAS can be interpreted as the Fermi level, with positive and negative changes for transition above and below the Fermi level, respectively.

The latest joint theoretical and experimental work on pure Ni [17] shows that the presence of a finite Hubbard on-site correlation term U, as calculated by time-dependent density functional theory, yields the typical XAS line shape [18, 19]. This explanation is also viable for antiferromagnetic materials, such as NiO [20]. Further progress in the exact understanding of the simultaneous demagnetization and XAS changes has been hindered by the lack of systematic studies of the fluence dependence of the transient changes and their fundamental relationship to the precise electronic properties of ferromagnetic systems.

Here, we report on a study of a CoPt alloy using femtosecond XAS and XMCD to probe the Co absorption $L_3$ edge at the self-amplified spontaneous emission (SASE) free-electron laser (FEL) FLASH in Hamburg [21, 22]. We observe the simultaneous dynamics of demagnetization and electronic changes in the valence bands of the compound close to the Fermi level. Comparison to previous measurments for CoPd indicate a more efficient demagnetization process for CoPt.

II. EXPERIMENT

We performed transient XAS and XMCD spectroscopy experiments in transmission geometry. The sample consisted of a 25nm thick $Co_{50}Pt_{50}$ alloy layer, prepared by magnetron sputtering at the Max Born Institute, Berlin, Germany. The sample was capped by 2 nm of MgO and separated from the supporting substrate by 2 nm of Ta. The transmission geometry of our experiments required a substrate transparent to X-rays, which in our case is a 200 nm thick SiN membrane. The opposite side of the membrane was covered by 100 nm of Al, acting as a heat sink. Samples were mounted on a Si chip with appropriate openings and transferred to the MUSIX experimental chamber [23]. Measurements were performed at room temperature. The in-plane magnetization of the sample was manipulated by an electromagnet, reaching a field sufficient to saturate the magnetization in-plane only (see the hysteresis in the right part of Fig. 1.). The XMCD effect is proportional to $\mathbf{k} \cdot \mathbf{M}$, where $\mathbf{k}$ is the X-ray beam propagation vector, and $\mathbf{M}$ is the magnetization. The X-ray incidence angle was set to 35° relative to the surface normal.

Our measurements were performed at beamline FL23 [24] of FLASH2, which covers the L-edges of the important $3d$ transition metal elements using the 3$^{rd}$ harmonic radiation of FLASH2 [23]. An overview of the experimental setup is shown in Fig. 1. We used circularly polarized soft X-rays from the new helical APPLE-III afterburner undulator [25]. This afterburner undulator has a magnet configuration such that its fundamental radiation amplifies the third harmonic radiation of the main undulators though with circular polarization. For XAS measurements, the photon energy was varied by simultaneous adjustment of the FLASH2 undulators, afterburner, and FL23 monochromator. FLASH provides photon pulses in pulse trains with a repetition rate of 10 Hz, here with 400 pulses within each train, and an intra-train repetition rate of 1 MHz. The beamline monochromator was operated in a single-diffraction grating mode to select the radiation at the L-edge resonances. The varied line spacing (VLS) grating provided an energy dispersion of 0.017 nm/mm from the high-energy grating (600 lines/mm central groove density). It is followed by an exit slit of 100 micrometer, leading to an energy bandwidth of about 800 meV at the Co $L_3$-edge of 780 eV. An adaptive Kirkpatrick-Baez (KB) focusing optics system focused the FEL beam onto the CoPt sample. We estimated the FEL pulse duration at the sample to be <80 fs FWHM by considering the expected FEL pulse length (<50 fs) and the stretching due to the monochromator grating (<30 fs). The total time resolution is lowered further by arrival time jitter between the FEL radiaton and the optical pump pulses. In order to normalize the intrinsic pulse-to-pulse SASE intensity fluctuations, we recorded the incoming $I_0$-intensity by splitting the beam after the monochromator slit using a transmission grating, measuring the pulse-resolved intensity of the gratings 1st diffraction order while the 0th order continues to the sample for absorption measurements [26, 27].

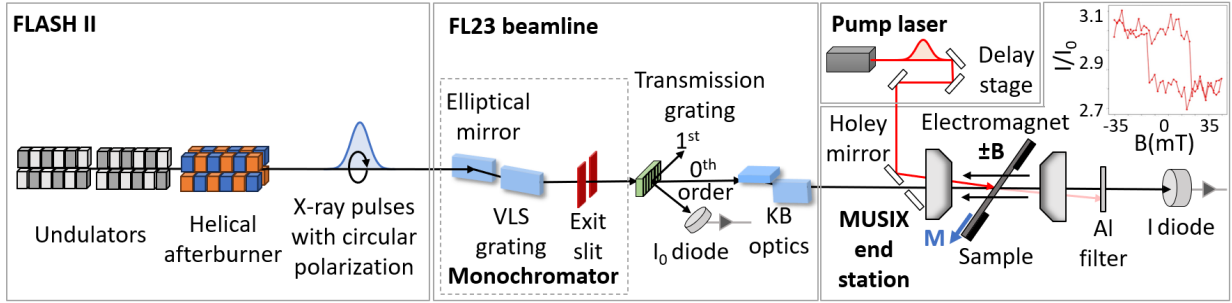

**Fig. 1. Schematic representation of the experimental setup.** X-rays with right circular polarization were produced at FLASH2 using a combination of planar undulators and an APPLE-III helical afterburner undulator. The X-ray beam transport and shaping were done in the FL23 beamline, consisting of a single VLS grating monochromator, an exit slit, an $I_0$ monitor, and refocusing optics. The monochromatic beam was split by a transmission grating, with a fast diode placed in one of the 1st-order diffracted beams, recording the incoming $I_0$ intensity, and the 0th-order was used to illuminate the sample. The absorption experiment was done in the MUSIX endstation in transmission geometry, where the sample was placed within an electromagnet, and the transmitted intensity, I, was monitored by a diode. By sweeping the magnetic field and monitoring the $I/I_0$ ratio, we obtained a magnetic hysteresis loop corresponding to the projection of the in-plane magnetization onto the X-ray beam direction. The final time-resolved experiments were done by coupling the pump laser into the MUSIX endstation. The corresponding time delay was set by a mechanical delay stage. Note that not all components are shown in Fig. 1 (for details, see [28, 29]).

The optical pump laser with a wavelength of 1030 nm delivered pulses of 65 fs FWHM duration at 100 kHz repetition rate and was coupled into the MUSIX chamber along with the FEL beam in a collinear fashion (see Fig. 1). The optical pump fluence was set to 4.2 mJ/cm$^2$. For the time-resolved measurements, we use the sequence of 40 FEL pulses within the pulse train, where every tenth pulse is accompanied by a pump pulse, and the other nine subsequent pulses remain unpumped. Pump-probe measurements allowed us to obtain time-delay traces at multiple fixed energies with high time resolution, as well as transient spectra over a wide range of photon energies.

The measurements consist of recording $I_0$ signals (upstream from the sample), and I signals (downstream) to normalize the intensity fluctuations of the individual pulses. A digitizer records each pulse, and these recorded data are integrated, averaging the pulses of each train. In postprocessing, we sort data based on train-by-train intensity and filter out too weak and too strong trains in I and $I_0$.

We also measured reference XAS and XMCD spectra at the BESSY II beamline PM3 on a sister sample to the one used at FLASH. Both samples were prepared at the same time under identical growth conditions. Measurements at BESSY II were performed using a monochromator energy resolution of 300 meV but otherwise the same conditions as for FLASH.

III. RESULTS:

A. Ground-state spectroscopy

Figures 2a and 2c compare the ground-state XAS and XMCD spectra for the Co $L_3$ edge obtained at FLASH (open green symbols) and BESSY (orange solid lines). L-edge XAS probes transitions from core $2p$ to unoccupied valence $3d$ states and is therefore sensitive to the state-resolved occupation relative to the Fermi level, $E_F$ [11]. L-edge XMCD measures the difference in occupation of the exchange-split valence states based on the significant spin-orbit splitting of the $2p_{3/2}$ and $2p_{1/2}$ core levels. In equilibrium, the valence levels are filled up to $E_F$ following the temperature-dependent Fermi distribution.

X-ray absorption was measured in transmission for circularly polarized X-rays for opposite sample mangetization directions (see Fig. 1). We extract the absorption as the logarithm of the ratio of the transmitted X-ray intensity after the sample, I, and the incoming intensity, $I_0$, recorded upstream of the sample (see Fig. 1). XAS in Fig. 2a is then the average of the two spectra for opposite magnetic field directions, while XMCD in Fig. 2c is given by their difference. Due to the high X-ray pulse intensity at FLASH, a small non-linearity of one of the detectors had to be taken into account for the XAS spectra. We used the BESSY II XAS measurements as a reference and normalized the FLASH spectra by a photon energy scan without a sample adjusted with a small intensity offset. We note that such corrections do not need to be applied for magnetic and time-resolved measurements since they are all based on calculating differences of individual XAS scans.

The BESSY XAS and XMCD spectra in Fig. 2 were convoluted by a Gaussian function to account for the differing spectral resolutions at BESSY II (300 meV) and FLASH (800 meV). This results in good agreement of the measured ground-state spectra obtained at FLASH and BESSY II.

B. Transient-state spectroscopy

Femtosecond laser excitation of a ferromagnetic metal results in the generation of electron-hole pairs around $E_F$. The hot, non-thermal electrons and holes interact with the spin and phonon subsystems, driving ultrafast demagnetization. The ultrafast redistribution of charge carriers leads to transient changes in the $2p$ to $3d$ transition probabilities for both spin-dependent and spin-independent terms of the matrix elements, resulting in characteristic lineshape changes in XAS and XMCD. By studying the temporal evolution of XAS and XMCD in the pump-probe experiment, we simultaneously obtain a quantitative picture of the evolution of the electronic and spin subsystems with sub-picosecond time resolution.

Figures 2b and 2c show the transient changes in the spectral profiles for XAS and XMCD, probing the $3d$ Co states at a delay of 500 fs after optical excitation (solid blue symbols). The transient differential X-ray absorption is obtained from the logarithm of the ratio of pumped and unpumped transmitted signals, normalized to their respective $I_0$ signals. In contrast, the unpumped and pumped XMCD signals are obtained as differences from the logarithm of the ratio of transmissions measured with opposite magnetic fields (i.e., opposite directions of **M**). The most pronounced effect is the rapid loss of magnetization, which manifests in a constant decrease of the XMCD signal across the whole $L_3$ absorption edge. We used the XMCD lineshape of the BESSY II data as a reference and determined the XMCD attenuation relative to the unpumped signal to $\Delta$XMCD = 24 ± 9 %. At the same pump-probe time delay, the difference in the XAS spectrum demonstrates a characteristic zero-crossing slightly below the XAS absorption maximum (Fig. 2b). This is in good agreement with previous measurments on CoPd multilayers [11] where a much stronger $\Delta$XAS change was observed, however, with the $\Delta$XAS zero-crossing and, thus, $E_F$ at similar energy positions relative to the CoPd XAS measurements. Using the $\Delta$XAS lineshape obtained in Ref.

[11] as a fit function (blue solid line in Fig. 2b) we determined the amplitude of the ΔXAS variation across $E_F$ to 0.7 ± 0.3 % relative to the XAS maximum in Fig. 2a.

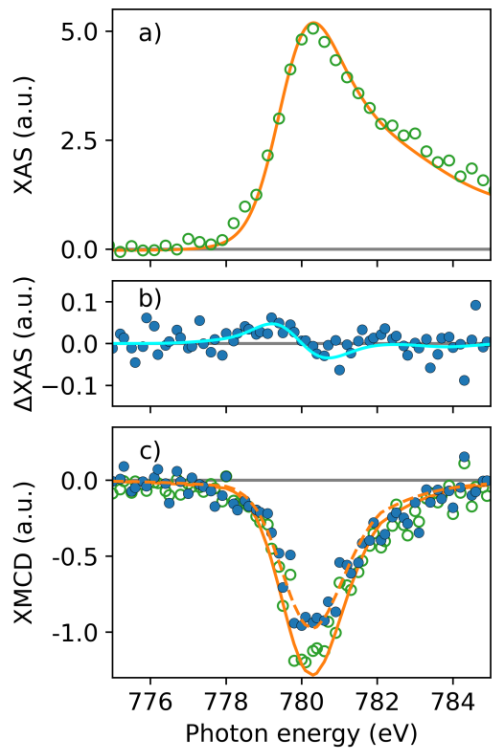

**Fig. 2. Ground-state and transient spectra across the $L_3$ edge.** (a) Ground state XAS spectrum of CoPt at the Co $L_3$-edge measured at FLASH (open green symbols) and BESSY II (solid orange line). (b) Transient differential spectrum ΔXAS (solid blue symbols), given by the difference between pumped and unpumped XAS and recorded at 500 fs pump-probe time delay. The blue solid line is a fit in amplitude of the ΔXAS spectrum from Ref. [11]. Relative to the XAS maximum in (a), the ΔXAS change in (b) exhibits a 0.7 ± 0.3%. (c) Ground-state (open green symbols) and transient XMCD spectra recorded at 500 fs delay (solid blue symbols). XMCD spectra taken at BESSY II are shown as solid and dashed orange lines. The XMCD spectrum at 500 fs is reduced by 24 ± 9 % compared to the unpumped signal.

C. Time-resolved dynamcis

Here, we investigate the temporal evolution of Co 3*d* electron and spin dynamics. Figs. 3a and 3b display the ΔXAS and ΔXMCD variations vs. pump-probe time delay with the X-ray energy tuned to the $L_3$ absorption maximum.

We observed slow drifts of the synchronization between pump-laser system and FLASH. This led to slow variations of the zero time-delay position over several hours as also observed in Ref. [11]. Therefore, multiple experimental runs were analyzed separately, yielding similar overall temporal variations but with larger noise levels. The slight time-zero, $t_0$, shifts were corrected by shifting to a common $t_0 = 0$ ps and averaging these runs following the approach in Ref. [11]. The averaged data are displayed in Fig. 3.

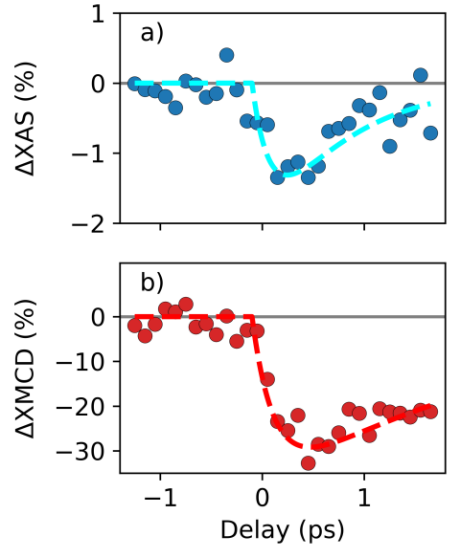

**Fig. 3. Time-resolved dynamics.** (a) Transient changes in XAS (blue solid symbols) and, (b) in XMCD (red solid symbols). The data have been recorded with the X-ray energy set to the maximum of the $L_3$ resonance and the unpumped values were subtracted. The dashed lines are double exponential fits, as described in the text.

We observe the expected ultrafast response in XAS and XMCD close to time-zero [9, 11]. Following this initial signal drop, recoveries at longer timescales are observed in Fig. 3. Following Refs. [9, 11] we fitted both data sets (solid lines in Figs. 3a and b) by a multiplicative double exponential with time constants, $\tau_D$ and $\tau_R$, describing decay and recovery, respectively. However, due to the relatively large noise level in the delay scans of Fig. 3 we kept the time constants fixed and adjusted only the amplitudes of the fit functions. This resulted in a variation of $\Delta$XAS = 1.3 ± 0.3 % and $\Delta$XMCD = 30 ± 3 % with respect to the unpumped data. The initial decay is $\tau_D = 0.3$ ps for the XMCD signal and therefore close to the expected decay time constant of about 100 fs [31, 32] convoluted by the estimated temporal resolution of 170 – 270 fs. We choose $\tau_D = 0.3$ ps also for the $\Delta$XAS dataset due to its larger statistical uncertainty. The recovery dynamics is different for ΔXAS with $\tau_{R,XAS}$ = 0.7 ps and for ΔXMCD with $\tau_{R,XMCD}$ = 2.5 ps. The different time constants are attributed to the timescale of electron-phonon equilibration as observed with XAS compared to the slower remagnetization (recovery of spin excitations) seen in the XMCD signal [9, 11, 30].

IV. DISCUSSION

The CoPt data shown in Figs. 2 and 3 represent the first time-resolved XAS and XMCD measurements at FLASH with the new helical afterburner undulator. Here we compare our data with previous results for CoPd obtained at LCLS using the DELTA undulator [11] which was the first such helical insertion device at an X-ray FEL [14]. CoPd was also previously studied at the BESSY II femtosecond slicing source [9]. The laser-induced demagnetization visible as a reduction of the XMCD signal is in all cases accompanied by a repopulation of valence electrons around the Fermi level.

We can compare these three experiments by extrapolating the measured demagnetization to the expected amount of demagnetization for the sample volume starting at the surface down to the optical pump laser penetration depth. In our sample, for the used pump fluence of 4.2 mJ/cm$^2$, we estimate a roughly 100% demagnetization in the near-surface region down to about 10 nm, the approximate laser penetration depth [11]. This is comparable to the measurement on a 15 nm CoPd film [9] where a 60 % demagnetization at 12 mJ/cm$^2$ pump fluence was reported. The CoPd sample in Ref. [11] had a thickness of 46 nm probed at normal X-ray incidence and showed a 60 % demagnetization at 35 mJ/cm$^2$ pump fluence. The much higher pump fluence used in this study [11] leads to a significant overheating in the surface layer by roughly twice the pump fluence necessary to completely demagnetize the material.

Turning now to the discussion of the laser-induced electron-hole dynamics we expect that this scales linearly with the optical pump fluence. This implies that the near surface layers in CoPd of Ref. [11] should show a ΔXAS change that is stronger than that for the more moderately pumped CoPd sample in Ref. [9]. Boeglin et al. report a transient ΔXAS signal of up to 14% measured about 0.8 eV below the XAS maximum [9]. The CoPd sample studied by Le Guyader et al. dispayed a 10 % ΔXAS signal at the same energy position [11]. If we take the larger film thickness into account this must correspond to a significantly larger ΔXAS change in the near-suface region.

Contrary tho both studies on CoPd our results for CoPt display an almost negligible ΔXAS change as seen in Figs. 2b and 3a. This is in line with the much smaller pump fluence used.

Our study therefore indicates that CoPt alloys can be more efficiently, i.e. at lower optical pump fluences, demagnetized than CoPd. This is in agreement with previous studies [31-35] that observed an enhanced demagnetization of Co/Pt multilayers relative to pure Co. $Co_{50}Pt_{50}$ and $Co_{50}Pd_{50}$ alloys have very similar magnetic ordering temperatures of 850 K and 950 K, respectively [36]. Since the measurements here and in the literature [9, 11] were performed at room temperature one would not expect a significantly different amount of demagnetization for similar pump conditions based on the magnetic ordering temperature alone. The enhanced demagnetization rather points to an increase of the spin-flip probabilities related to the size of valence band spin-orbit coupling as has been observed in Co/Pt multilayers [31]. This spin-orbit coupling enhancement also leads to the well-known increase of Co orbital magnetic moments as can be probed with XMCD [37].

V. SUMMARY AND CONCLUSIONS

In this work, we have reported the first time-resolved X-ray magnetic circular dichroism measurements using a new helical afterburner undulator at FLASH. Contrary to previous studies [31-35] that have been performed at the Co M-edges or in the optical spectral region our study performed at the Co L-edges allows the unambiguous separation of the laser-induced electron repopulation and magnetization dynamics. We showed that CoPt alloys display the characteristic transient XAS lineshape upon demagnetization as observed in previous studies for CoPd at LCLS [11]. We found that CoPt displays a similar degree of ultrafast demagnetization as CoPd, however, at significantly reduced optical pump fluences. We attribute this efficient demagnetization channel to the stronger spin-orbit coupling in Pt compared to Pd leading to very efficient spin-flip processes in CoPt alloys driving ultrafast demagnetization.

ACKNOWLEDGMENTS

We acknowledge DESY (Hamburg, Germany), a member of the Helmholtz Association HGF, for the provision of experimental facilities. Parts of this research were carried out at the beamline

FL23 at FLASH2. Beamtime was allocated for proposals F-20211728 EC and F-20220710. Work at Uppsala University was supported by the Swedish Research Council (VR) and the K. and A. Wallenberg Foundation. Work at Freie Universität Berlin, Max Born Institute, and Helmholtz-Zentrum Berlin was funded by the German Research Foundation (DFG) through TRR277 (Project-ID 328545488), Projects A01, A02 and A03. M.W. acknowledges funding from Federal Ministry for Education and Research (BMBF), project Spinflash (Project-ID 05K22KE2).

AUTHOR DECLARATIONS

Conflict of Interest: The authors have no conflicts to disclose.

DATA AVAILABILITY

The data that support the findings of this study are available from the corresponding author.

AUTHOR CONTRIBUTIONS

**M. Pavelka**: Formal analysis (equal); Investigation (equal); Methodology (equal); Software (equal); Visualization (equal); Writing – original draft (lead); Writing – review & editing (equal). **S. Marotzke**: Formal analysis (equal); Investigation (equal); Methodology (equal); Software (equal); Visualization (equal); Writing – original draft (equal); Writing – review & editing (equal). **R.-P. Wang**: Formal analysis (equal); Investigation (equal); Methodology (equal); Software (equal); Writing – review & editing (equal). **M. Elhanoty**: Methodology (equal); Writing – review & editing (equal). **G. Brenner**: Conceptualization (equal); Investigation (equal); Methodology (equal); Project administration (equal); Visualization (equal); Writing – original draft (equal); Writing – review & editing (equal). **S. Dziarzhytski**: Investigation (equal); Methodology (equal); Writing – review & editing (equal). **S. Jana**: Formal analysis (equal); Investigation (equal); Writing – review & editing (equal). **D. Engel**: Resources (lead); Writing – review & editing (equal). **C. von Korff Schmising**: Conceptualization (equal); Investigation (equal); Resources (equal); Writing – review & editing (equal). **D. Gupta**: Formal analysis (equal); Writing – review & editing (equal). **I. Vaskivskyi**: Formal analysis (equal); Investigation (equal); Writing – review & editing (equal). **T. Amrhein**: Formal analysis (equal); Investigation (equal); Writing –review & editing (equal). **N. Thielemann-Kühn**: Formal analysis (equal); Investigation (equal); Writing – review & editing (equal). **M. Weinelt**: Funding acquisition (equal); Investigation (equal); Writing – review & editing (equal). **R. Knut**: Investigation (equal); Writing – review & editing (equal). **J. Rönsch-Schulenberg**: Methodology (equal); Writing – review & editing (equal). **E. Schneidmiller**: Investigation (equal); Methodology (equal); Writing – original draft (equal); Writing – review & editing (equal). **C. Schüßler-Langeheine**: Investigation (equal); Methodology (equal); Writing – review & editing (equal). **M. Beye**: Conceptualization (equal); Formal analysis (equal); Investigation (equal). Methodology (equal); Project administration (equal); Software (equal); Writing – review & editing (equal). **N. Pontius**: Investigation (equal); Methodology (equal); Writing – review & editing (equal). **O. Grånäs**: Conceptualization (equal); Funding acquisition (equal); Investigation (equal); Methodology. **H. A. Dürr**: Conceptualization (lead); Funding acquisition (equal); Investigation (equal); Methodology (equal); Project administration (equal); Supervision (equal); Visualization (equal); Writing – original draft (equal); Writing – review & editing (equal).